\documentclass[]{spie}

\usepackage[]{graphicx}

\title{\Large\bf Quantum Diagrams and Quantum Networks}

\author{Louis H. Kauffman\supit{a} and Samuel J. Lomonaco Jr.2\supit{b}
\skiplinehalf
\supit{a} Department of Mathematics, Statistics and Computer Science  
(m/c 249), 851 South Morgan Street, University of Illinois at Chicago,
Chicago, Illinois 60607-7045, USA \\
\supit{b} Department of Computer Science and Electrical Engineering, University of
Maryland Baltimore County, 1000 Hilltop Circle, Baltimore, MD 21250, USA}
 
\authorinfo{Further author information: L.H.K.  E-mail: kauffman@uic.edu,
S.J.L. Jr.: E-mail: lomonaco@umbc.edu}
 
\begin{document} 
  \maketitle 

%%%%%%%%%%%%%%%%%%%%%%%%%%%%%%%%%%%%%%%%%%%%%%%%%%%%%%%%%%%%% 
\begin{abstract} This paper explores how diagrams of quantum processes can be used for modeling and for quantum epistemology.
\end{abstract}


%%%%%%%%%%%%%%%%%%%%%%%%%%%%%%%%%%%%%%%%%%%%%%%%%%%%%%%%%%%%%
\section{INTRODUCTION}
This paper is an introduction to diagrammatic methods for representing quantum processes and quantum computing. We review basic notions for quantum information and quantum computing in Sections 2 and 3. In Section 3, we discuss topological diagrams and some issues about using category theory in representing quantum computing and teleportation. In Section 4 we analyze very carefully the diagrammatic meaning of the usual representation of the Mach-Zehnder interferometer, and we show how it can be generalized to associate to each composition of unitary transformations a ``laboratory 
thought experiment diagram" such that particles moving though the many alternate paths in this diagram will mimic the quantum process represented by the composition of unitary transformations. This is a finite dimensional way to think about the Feynman Path Integral. We call our representation result the
{\it Path Theorem.} In Section 5 we go back to the basics of networks and matrices and show how elements of quantum measurement can be represented with network diagrams.
\bigbreak

\section{Quantum Mechanics and Quantum Computation}
We shall indicate the basic principles of quantum mechanics.  The quantum information context 
encapsulates a concise model of quantum theory:
\bigbreak

{\em The initial state of a quantum process is a vector $|v \rangle$ in a complex vector space $H.$
Measurement returns basis elements $\beta$ of $H$ with probability 

$$|\langle \beta \,|v \rangle |^{2}/\langle v \,|v \rangle$$

\noindent where $\langle v \,|w \rangle = v^{\dagger}w$ with $v^{\dagger}$ the conjugate transpose of $v.$
A physical process occurs in steps $|v\rangle \longrightarrow U\,|v \rangle = |Uv \rangle $ where $U$ is a unitary linear transformation. Measurement and quantum process are kept separate in the model.
\bigbreak

Note that since $\langle Uv \,|Uw \rangle = \langle v \,|U^{\dagger}U |w \rangle = \langle v \,|w \rangle = $ when $U$ is unitary, it follows that probability
is preserved in the  course of a quantum process.  }
\bigbreak

One of the details required for any specific quantum problem is the nature of the unitary 
evolution.  This is specified by knowing appropriate information about the classical physics that 
supports the phenomena. This information is used to choose an appropriate Hamiltonian through which the 
unitary operator is constructed via a correspondence principle that replaces classical variables with appropriate quantum
operators. (In the path integral approach one needs a Langrangian to construct the action on which the path
integral is based.) One needs to know certain aspects of classical physics to 
solve any specific quantum problem.  
\bigbreak

A key concept in the quantum information viewpoint is the notion of the superposition of states.
If a quantum system has two  distinct states $|v \rangle$ and $|w \rangle,$ then it has infinitely many states of the form
$a|v \rangle + b|w \rangle$ where $a$ and $b$ are complex numbers taken up to a common multiple. States are ``really" 
in the projective space associated with $H.$ There is only one superposition of a single state $|v \rangle$ with 
itself. On the other hand, it is most convenient to regard the states $|v \rangle$ and $|w \rangle$ as vectors in a vector space.
We than take it as part of the procedure of dealing with states to normalize them to unit length. Once again, the superposition of a state with itself is
again itself.
\bigbreak

Dirac introduced the ``bra -(c)-ket" notation $\langle A\,|B \rangle = A^{\dagger}B$ for the inner product of complex vectors $A,B \in H$.
He also separated the parts of the bracket into the {\em bra} $\langle  A\,|$ and the {\em ket} $|B \rangle.$ Thus

$$\langle A\,|B \rangle = \langle A\,|\,\,|B \rangle$$

\noindent In this interpretation,
the ket $|B \rangle$ is identified with the vector $B \in H$, while the bra $<A\,|$ is regarded as the element dual to $A$ in the 
dual space $H^*$. The dual element to $A$ corresponds to the conjugate transpose $A^{\dagger}$ of the vector $A$, and the inner product is 
expressed in conventional language by the matrix product $A^{\dagger}B$ (which is a scalar since $B$ is a column vector). Having separated the bra and the ket, Dirac can write the
``ket-bra"  $|A \rangle \langle B\,| = AB^{\dagger}.$ In conventional notation, the ket-bra is a matrix, not a scalar, and we have the following formula for
the  square of $P = |A \rangle \langle B\,|:$

$$P^{2} =  |A \rangle \langle B\,| |A \rangle \langle B\,| = A(B^{\dagger}A)B^{\dagger} = (B^{\dagger}A)AB^{\dagger} = \langle B\,|A \rangle P.$$

\noindent The standard example is a ket-bra $P = |A\,\rangle \langle A|$ where $\langle A\,|A \rangle =1$ so that $P^2 = P.$  Then $P$ is a projection
matrix,  projecting to the subspace of $H$ that is spanned by the vector $|A \rangle$. In fact, for any vector $|B \rangle$ we have 

$$P|B \rangle = |A \rangle \langle A\,|\,|B \rangle =  |A \rangle \langle A\,|B \rangle = \langle A\,|B \rangle |A \rangle .$$

\noindent If $\{|C_{1} \rangle, |C_{2} \rangle , \cdots |C_{n} \rangle \}$ is an orthonormal basis for $H$, and $$P_{i} = |C_{i} \,\rangle \langle C_{i}|,$$
\noindent then for any vector $|A \rangle $ we have

$$|A \rangle = \langle C_{1}\,|A \rangle |C_{1} \rangle + \cdots + \langle C_{n}\,|A \rangle |C_{n} \rangle .$$

\noindent Hence 

$$\langle B\,|A \rangle = \langle B\,|C_{1} \rangle \langle C_{1}\,|A \rangle + \cdots + \langle B\,|C_{n} \rangle \langle C_{n}\,|A \rangle $$

One wants the amplitude for starting in state $|A \rangle $ and ending in state $|B \rangle .$ The probability for this event is equal to $|\langle B\,|A \rangle |^{2}$. This can be refined if we have more knowledge. 
If the intermediate states $|C_{i} \rangle $ are a complete set of orthonormal alternatives then we
can assume that 
$\langle C_{i}\,|C_{i} \rangle  = 1$ for each $i$ and that $\Sigma_{i} |C_{i} \rangle \langle C_{i}| = 1.$  This identity now corresponds to the fact that
$1$ is the sum of the amplitudes of an arbitrary state projected into one of these intermediate states.
\bigbreak

If there are intermediate states between the intermediate states this formulation can be continued
until one is summing over all possible paths from $A$ to $B.$ This becomes the path integral expression 
for the amplitude $\langle B|A \rangle .$
\bigbreak

\subsection{What is a Quantum Computer?}

A {\it quantum computer} is a composition $U$ of unitary transformations, together with an initial state and a choice of measurement
basis. One runs the computer by repeatedly initializing it, and then measuring the result of applying the unitary transformation $U$ to the initial state.
The results of these measurements are then analyzed for the desired information that the computer was set to determine.  
\bigbreak

Let $H$ be a given finite dimensional vector space over the complex numbers $C.$ Let 
$$\{ W_{0}, W_{1},..., W_{n} \}$$ be an
orthonormal basis for $H$ so that with $|i \rangle := |W_{i} \rangle $ denoting $W_{i}$ and $\langle i|$ denoting the conjugate transpose of $|i \rangle $,
we have
$$\langle i|j \rangle = \delta_{ij}$$
\noindent where $\delta_{ij}$ denotes the Kronecker delta (equal to one when its indices are equal to one another, and equal
to zero otherwise). Given a vector $v$ in $H$ let $|v|^{2} := \langle v|v \rangle .$ Note that $\langle i|v$ is the $i$-th coordinate of $v.$ 
\vspace{3mm}

\noindent An {\em measurement of $v$} returns one of the coordinates $|i \rangle $
of $v$ with probability $|\langle i|v|^{2}.$ This model of measurement is a simple instance of the situation with a quantum
mechanical system that is in a mixed state until it is observed. The result of observation is to put the system into one of
the basis states. 
\vspace{3mm}

When the dimension of the space $H$ is two ($n=1$), a vector in the space is called a {\em qubit}. A qubit represents one
quantum of binary information. On measurement, one obtains either the ket $|0 \rangle $ or the ket $|1 \rangle $. This constitutes the 
binary distinction that is inherent in a qubit.  Note however that the information obtained is probabilistic.  If the qubit is
$$| \psi \rangle = \alpha |0 \rangle + \beta \ |1 \rangle ,$$ \noindent then the ket $|0 \rangle $ is observed with probability $|\alpha|^{2}$, and the ket
$|1 \rangle $ is observed with probability $|\beta|^{2}.$  In speaking of an idealized quantum computer, we do not specify the nature
of measurement process beyond these probability postulates.
\vspace{3mm}
 
In the case of general dimension $n+1$ of the space $H$, we will call the vectors in $H$
{\em qunits}. It is quite common to use spaces $H$ that are tensor products of two-dimensional spaces (so that all computations 
are expressed in terms of qubits) but this is not necessary in principle. One can start with a given space, and later work out
factorizations into qubit transformations.
\vspace{3mm}

A {\em quantum computation} consists in the application of a unitary
transformation $U$ to an initial qunit 
$$|\psi \rangle  = a_{0}|0 \rangle + ... + a_{n}|n \rangle $$  with $|\psi|^{2}=1$, plus an
measurement of
$U|\psi\rangle .$ A measurement of $U|\psi\rangle$ returns the ket $|i \rangle $ with probability $|\langle i|U|\psi\rangle|^{2}$. In particular, if we start the computer
in the state $|i \rangle $, then the probability that it will return the state $|j \rangle $ is $|\langle j|U|i \rangle |^{2}.$

\vspace{3mm} It is the necessity for writing a given computation in terms of unitary transformations, and the probabilistic
nature of the result that characterizes quantum computation. Such computation could be carried out by an idealized quantum
mechanical system.  
\vspace{3mm}

\begin{figure}
     \begin{center}
     \begin{tabular}{c}
     \includegraphics[width=8cm]{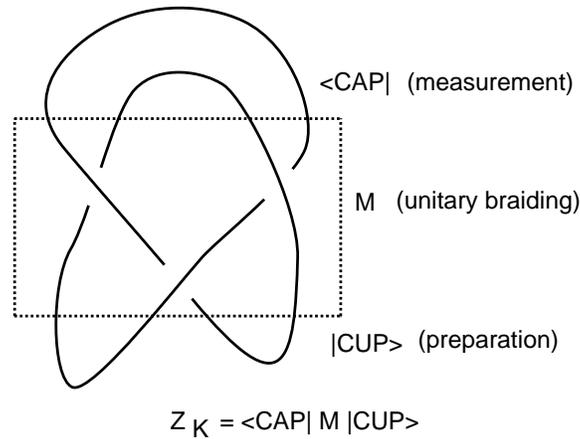}
     \end{tabular}
     \caption{\bf A Knot Computer}
     \label{knotcomp}
\end{center}
\end{figure}

We end this section with two diagrams that show some of the ideas behind quantum computation and also illustrate the diagrammatic issues of this paper. In Figure~\ref{knotcomp} we show the diagram for a quantum computer based on knot theoretic structure. In this diagram the crossings in the knot represent
unitary transformations that have topological properties \cite{BG,CKL,QCJP,AnyonicTop}. The cups and the caps denote preparations and measurements respectively. The point we make about this sort of diagram is that it promotes thinking back and forth between topology and quantum computing. One can use the language of category theory \cite{C2} to clarify these interdiscipinary issues. The language of category theory is particularly useful when there is a geometric/topological object, such as a knot, that is subject to being divided into parts that represent sub-processes of a larger process. Our attitude toward the use of category theory
in quantum computing is based on this notion of decomposition. When a process is specified by a space 
or a graph, there may be many ways to factor it into subprocesses. An appropriate choice of category and generators of that category (here the crossings, the cups and the caps) can act to organize and
clarify the structure of the situation and the relationship between the disciplines that are under discussion.

\begin{figure}
     \begin{center}
     \begin{tabular}{c}
     \includegraphics[width=8cm]{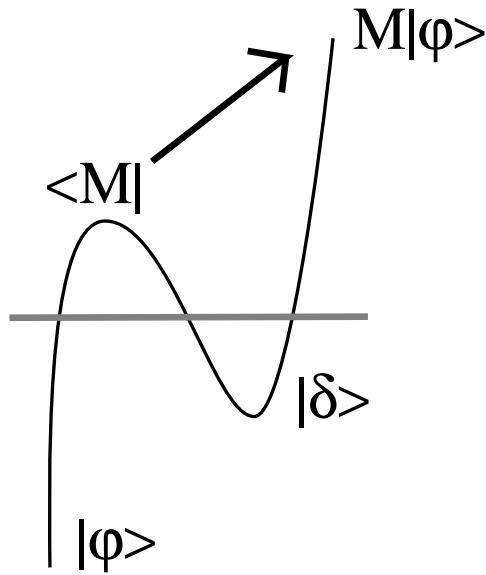}
     \end{tabular}
     \caption{\bf Teleportation}
     \label{teleport}
\end{center}
\end{figure}

Figure~\ref{teleport} illustrates a diagram for teleportation \cite{Teleport,TEQE,Spie,C1}. Here the cap $\langle M|$
represents a measurement, while the cup $|\delta \rangle$ represents an an entangled (EPR) state.
The direction for process and measurement proceeds from the bottom of the diagram to the top, through
which the initial state $|\phi\rangle$ is transformed to $M |\phi\rangle.$ Here $M$ denotes a unitary 
transformation that is naturally associated with the measurement. On the other hand, from a graphical point of view, the composition of transformations is {\it mathematically equivalent} to what one finds on traversing the one-dimensional arc of the curve of the cap followed by the cup. This corresponds to the composition of the matrices $M$ and $\delta$ (an identity matrix in this case) and shows directly how the state $|\phi\rangle$ was transformed to $M|\phi\rangle.$ Underneath any particular point of view about category, direction (or time in the physical interpretation) is the structure embodied in compositions of transformations. These are directly represented by the diagrams. Multiple intepretations of the diagrams furnish a multiplicity of viewpoints for the problem at hand.
\bigbreak

\section{Half-Silvered MIrrors, a Mach-Zehnder Interferometer and the Path Theorem}
In this section we give examples of diagramming quantum processes and we prove that any finite dimensional quantum process can be configured as a thought experiement that generalizes the 
diagrammatics for the Mach-Zehnder Interferometer.
\bigbreak

\begin{figure}
     \begin{center}
     \begin{tabular}{c}
     \includegraphics[width=8cm]{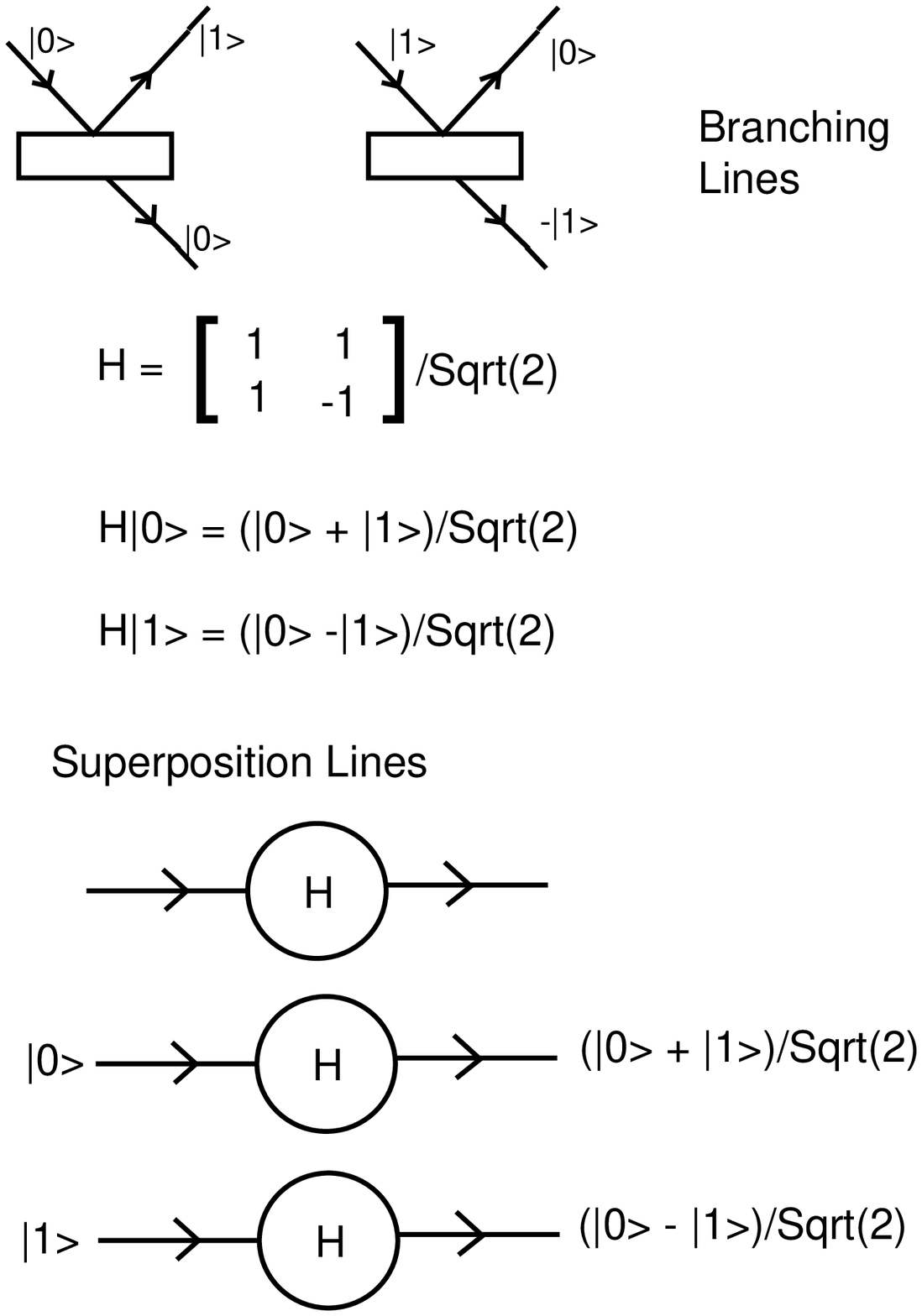}
     \end{tabular}
     \caption{\bf Hadamard as Half-Silvered Mirror}
     \label{hadamard}
\end{center}
\end{figure}

View Figure~\ref{hadamard}. In this figure we indicate a matrix $H$ that acts on a qubit space via the formulas 
$$H |0 \rangle = \frac{1}{\sqrt{2}}(|0 \rangle + |1 \rangle),$$
$$H |1 \rangle = \frac{1}{\sqrt{2}}(|0 \rangle - |1 \rangle).$$
This is a unitary transformation. In the same figure we indicate a diagram that can be interpreted as a half-silvered mirror that admits or transmits a particle that is labeled either as $|0\rangle$ or as
$|1\rangle.$ We shall refer to these as the $0$ and $1$ states of the particle. The rules by which the mirror operates are that when it transmits a particle, it does not change the state of the particle, but it may change the phase. In this case when the mirror transmits a $1$ it changes the phase corresponding to the negative sign in the second formula. But this mirror is supposed to be a ``quantum mirror," and so if a state of $0$ or $1$ enters the mirror, what ``leaves" the mirror is a superposition of $0$ and $1$ along the two possible exit paths of transmission and reflection. As the reader can see, the diagram we have used for the half-silvered mirror is a combination of what one might diagram from a laboratory situation, combined with the principles of the quantum model. The diagram includes information that mimics the laboratory, with the understanding that the mirror action means that detection in a certain direction indicates reflection, while detection in another direction indicates transmission. 
\bigbreak

In the same Figure~\ref{hadamard} we have indicated a more abstract diagram for the Hadamard matrix that just indicates an input or preparation line and shows the linear superposition on the output line.
This sort of diagram can be used to indicate quantum processes quite independently of extra laboratory descriptions. 
\bigbreak

\begin{figure}
     \begin{center}
     \begin{tabular}{c}
     \includegraphics[width=8cm]{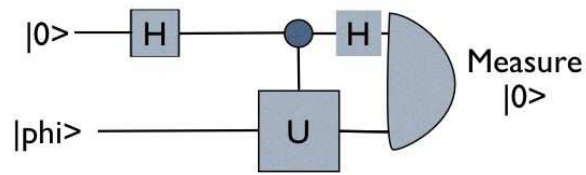}
     \end{tabular}
     \caption{\bf Hadamard Test}
     \label{hadamardtest}
\end{center}
\end{figure}

In Figure~\ref{hadamardtest} we show a diagram for a quantum computer, called the 
{\it Hadamard Test} that can be used to find matrix entries of the form $\langle \psi| U |\psi \rangle.$ The probability of observing $|0\rangle$ at the output of the test can be calculated to be
$\frac{1}{2} + \frac{1}{2}Re\langle \psi|U|\psi\rangle.$ This means that repeated use of the Test  shown here will yield approximate values for $\langle \psi|U|\psi\rangle.$ An insertion of an appropriate phase in the network, allows one to obtain approximations to the imaginary part. The language of this sort of 
diagrmmatic is quite independent of any laboratory specifics. The input lines are in parallel and indicate 
an algebraic tensor product of the input to the unitary transformation $U$ and an external line above it that carries a single qubit. To the right of the inputs is a trivalent node and connection of the single qubit line with the unitary $U,$ indicating that $U$ has been extended to a controlled U, where a $|0\rangle$ on the single qubit line replaces $U$ by an identity transformation and a $|1\rangle$ on the line leaves $U$ operative. The controlled $U$ is itself a unitary operation and the diagram as a whole represents the composition of unitary operations. This is a typical example of a 
``quantum circuit diagram". We shall return to the general form of this kind of diagram in the next section.
\bigbreak

\begin{figure}
     \begin{center}
     \begin{tabular}{c}
     \includegraphics[width=8cm]{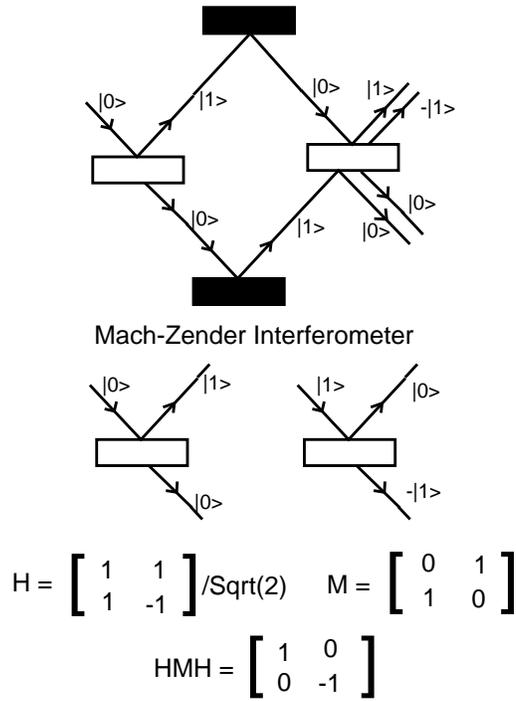}
     \end{tabular}
     \caption{\bf A Mach-Zehnder Interferometer}
     \label{machZehnder}
\end{center}
\end{figure}

In Figure~\ref{machZehnder} we have indicated a quantum apparatus made from half-silvered and ordinary mirrors. This diagram is a spatial composition of the half-silvered mirror diagram of Figure~\ref{hadamard} and another mirror diagram (indicated with dark shading) and matrix $M$ where
$$M |0 \rangle = |1\rangle,$$
$$M |1 \rangle =|1\rangle .$$
Thus, as a mirror, $M$ only reflects and changes type $0$ to type $1$ and vice-versa. We can continue to use the idea of a particle ``reflected" by this mirror, but it is understood that it just as well reflects a linear superposition as in the formula
$M(a |0\rangle + b|1\rangle) = a |1\rangle + b|0\rangle.$ Each mirror represents a unitary transformation
in a sequence of unitary transformations that move from the preparation to the detection diagonally from left in the diagram to the right. The diagram has been configured so that the reader can imagine particles moving through it in an imaginary laboratory. All four paths that the initial $|0\rangle$ particle can take 
through the apparatus have been indicated on the diagram. The reader will see that a ``destructive 
interference" occurs in the reflection output from the second half-silvered mirror, while a ``reinforcement" occurs in the transmission output from the second mirror. Thus we conclude, thinking of summing over all the paths, that a $|0\rangle$ qubit input will result in a $|0\rangle$ qubit output. This ``sum over paths" account of the quantum process for the Mach-Zhender interferometer is exactly coincident with the abstract process corresponding to the composition $HMH.$ What is the source of this coincidence?
\bigbreak

The conincidence is no coindidence. There is a $1-1$ correspondence between the indicated paths for a particle in the diagram for the interferometer and the product terms in the the matrix $HMH$ as it is 
applied to $|0\rangle$ (or to any other qubit as preparation). The key to this observaton is the definition of matrix multiplication: $$(AB)_{ij} = \Sigma_{k} A_{ik}B_{kj}.$$ With respect to a chosen basis the matrix element $A_{rs}$ represents the amplitude for a transition from state $r$ to state$s$ and so the matrix
entry $(AB)_{ij}$ is given by the sum over all products of amplitudes across all possibilities for state transition. In the case of the interferometer, the laboratory diagram is designed to encompass all these possibilities. But there is an important observation to make about this diagram. {\it The diagram contains two copies of the middle mirror $M.$} From the point of view of the quantum model we are using 
{\it these two copies of $M$ are identical in every way, in fact they are indistinguishable}. If an observer made a measurement to distinguish these two mirrors in the course of our laboratory thought experiment, paths in the process would be disturbed and we would not get the same result as 
$HMH |0\rangle.$ 
\bigbreak

\begin{figure}
     \begin{center}
     \begin{tabular}{c}
     \includegraphics[width=8cm]{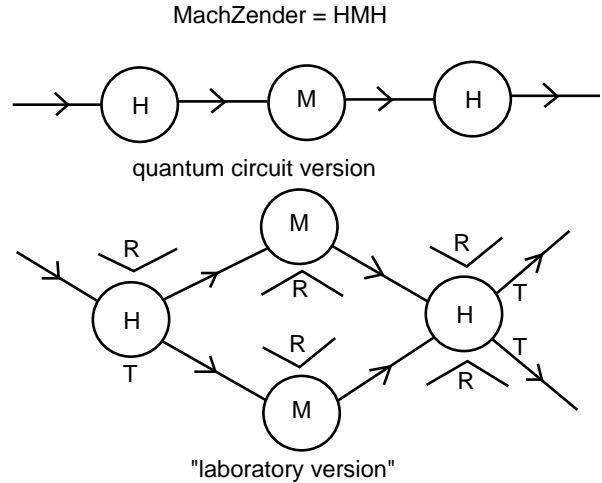}
     \end{tabular}
     \caption{\bf A Mach-Zehnder Circuit and Laboratory Path Gadget}
     \label{machcircuit}
\end{center}
\end{figure}

The laboratory diagram for the Mach-Zehnder interferometer contains features that are specific to interpreting $H$ and $M$ as mirrors. Thus we have set up the rules so that qubits will traverse certain lines when transmitted and other lines when reflected. In Figure~\ref{machcircuit} we have shown how the laboratory circuit we have described is unfolded from the straightline quantum circuit for the same process. In this diagram, we give the output lines roles according to reflection and transmission. 
\bigbreak

\begin{figure}
     \begin{center}
     \begin{tabular}{c}
     \includegraphics[width=8cm]{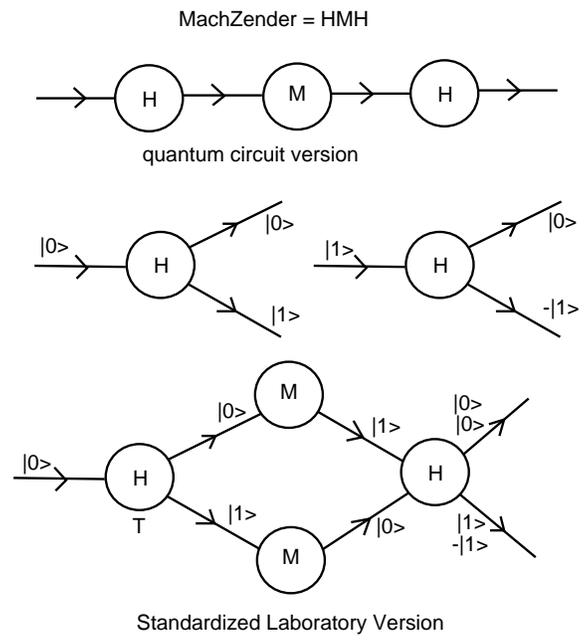}
     \end{tabular}
     \caption{\bf A Mach-Zehnder Circuit in Standardized Laboratory Form}
     \label{machcircuit1}
\end{center}
\end{figure}

However, a simpler convention can be taken for a more universal translation of compositions of unitary matrices to laboratory diagrams. In Figure~\ref{machcircuit1} we adopt the principle that the upper line on the diagram for $H$ indicates the $|0\rangle$ part of the output qubit, while the lower line indicates the
$|1\rangle$ part of the output qubit. In diagramming $M$ we make no change, since $M$ just interchanges
$|0\rangle$ and $|1\rangle.$ In this Figure we indicate the paths, and the reader can see easily that 
for the preparation $|0\rangle,$ there is destructive inference leading to only the output $|1\rangle.$
In this case it is easy to see the $1-1$ correspondence between the path summation and the matrix multiplication. Of course this laboratory diagram is less physical since we do not imagine that we can make a device that sorts the $0$ and $1$ states without performing a measurement! The mechanics of matrix multiplication provide the system of paths and give us a story about the branching nature of the quantum process. In order to be quantum mechanically accurate about such diagrams we cannot assume that any of the paths are independently observed. Only a quantum measurement is made at the end of the process.
\bigbreak

\begin{figure}
     \begin{center}
     \begin{tabular}{c}
     \includegraphics[width=8cm]{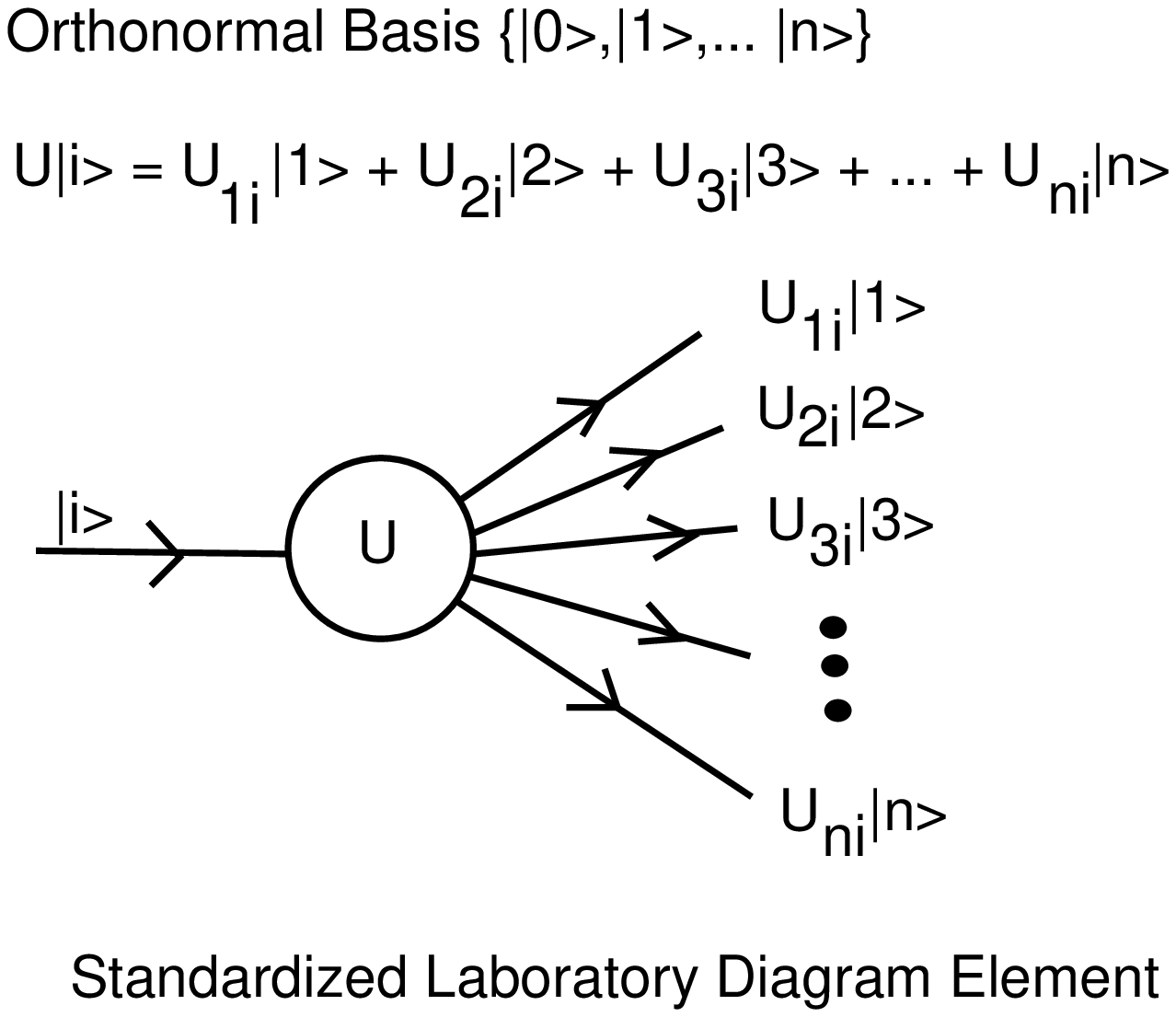}
     \end{tabular}
     \caption{\bf The General Standardized Laboratory Form for a Unitary Element}
     \label{node}
\end{center}
\end{figure}

In Figure~\ref{node} we give the full generalization of a {\it laboratory diagram node for a unitary transformation}. Here we assume that we work with a finite dimensional complex vector space $W$ with 
orthonormal basis ${\cal B}= \{|0\rangle, |1\rangle, \cdots |n\rangle\}$ and unitary transformation 
$U:W \longrightarrow W$ with matrix formula
$$U|i\rangle = \Sigma_{k=0}^{n} U_{ki}|k\rangle.$$

The multiplicity of lines emanating from the node represents the superposition of states in the sum above, {\it and} one can tell the story that an ``$|i\rangle$ particle " enters the $U$-device on the left and is  transformed into one of the outputs in each of a set of parallel universes with the indicated complex number weighting. If we then plug a number of these devices into one another, we get a generalized 
Mach-Zehnder interferometer whose action can be described by the summation over all these weighted 
paths each in its own parallel universe. Of course upon measurement at the end of the process, we invoke the quantum postulate and find that the path summations have conspired to give us the correct 
answers. We have proved the following 

\noindent {\bf Path Theorem.} {\it Every finite composition of unitary matrices can be represented by a ``quantum laboratory diagram'' such that the paths through the diagram are in $1-1$ correspondence with 
the products that are summed in the formation of the matrix product. Thus, one can construct a thought
experiment correpsonding to the composition of unitary matrices where particles move along a garden of forking paths, and the final amplitude for a given state is obtained by summing over the (complex number) contributions of each path.}
\bigbreak

\begin{figure}
     \begin{center}
     \begin{tabular}{c}
     \includegraphics[width=8cm]{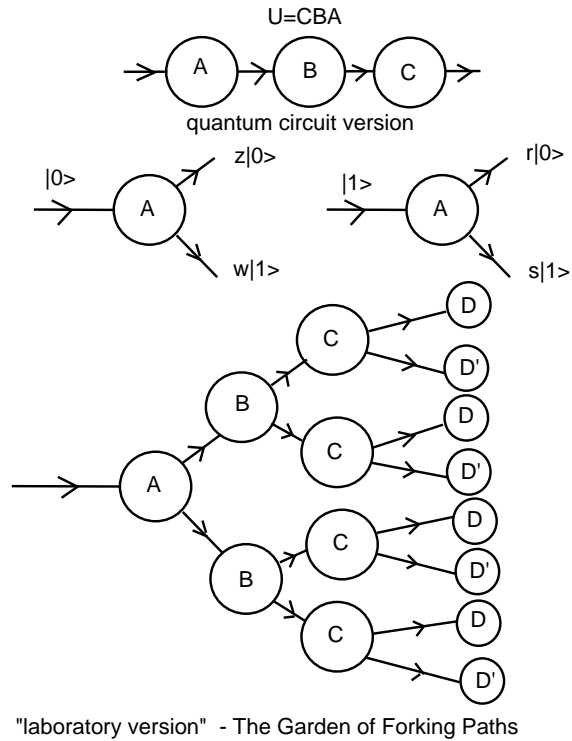}
     \end{tabular}
     \caption{\bf Unitary Composition as Garden of Forking Paths}
     \label{unit}
\end{center}
\end{figure}

In Figure~\ref{unit} we illustrate such a garden of forking paths for a composition of unitaries $A$, $B$ and $C$. Here the underlying space is a qubit space of dimension two, and so there are eight paths.
The diagrams labeled $D$ and $D'$ represent the two possibilities for detection of the superposition at the end of the process. One will either detect $|0\rangle$ and take the summation for all paths ending in 
$D$, or one will detect $|1\rangle$ and take the summation for all paths ending in $D'.$ These summations compute the amplitudes for the corresponding observations. 
\bigbreak

Note that in the composition of unitary transformations one must say that all paths occur at once. The paths do not happen in a multiplicity of worlds. The process happens in the Hilbert space for the quantum situation. There is no multiplicity of worlds.
But in the path-space interpretation, one is led to think of many different paths and the possibility that 
each path is traversed in a different world. It is clear from our diagrammatic constructions that the many worlds are a convenient fiction to enable one to imagine parallel paths when there are, in actuality, no paths at all.
\bigbreak

The Path Theorem is a finite dimensional analogue of the Feynman path integral. The formulation we have given raises questions about the relationship of the space of observations and spaces or graphs that may be standardly associated with a quantum process. We began this section with a version of the Mach-Zehnder interferometer whose geometry was, as a thought experiment, realistic in terms of the laboratory interpretation. In proving the Path Theorem we idealized each circuit element, removing the intepretation of {\it where} any given ``particle" would be detected. This meant that each circuit element comes equipped with a set of output lines corresponding abstractly to the question - where does one find a given particle? With the graphical geometry of these lines, concatenations of elements produce path possibilities, just as in the Mach-Zehnder interferometer, and hence a thought experiment for each quantum process. In a sense we are letting the quantum process {\it itself} produce a space of paths for its own elucidation. We intend to explore further the possibility that spaces that arise in physical situations can be construed as spaces that are fitting to their underlying quantum processes.
\bigbreak

\section{Diagrammatic Methods for Quantum Measurement}
The point of view of this paper is based on diagrammatic conventions for matrix multiplication and tensor composition.
The purpose of this section is to describe these conventions and to show how they are used in our work. We take diagrams to represent matrices and 
products or concatenations of matrices. In this way a complex network diagram can represent a contraction of a collection of multi-indexed matrices, and so may 
represent a quantum state or a quantum amplitude. We regard each graph as both a possible holder for matrices, and hence as a vehicle for such a computation, 
{\em and} 
as a combinatorial structure. As a combinatorial structure the graph can be modified. An edge can be removed. A node can be inserted. Such modifications
can be interpreted in terms of quantum preparation and measurement. One can then take the graph as a miniature ``world" upon which such operations are performed.
Since the graphs can also represent topological structures, this approach leads to a way to interface topology with the quantum mechanics.
The diagrammatics in this section should be compared with work of Roger Penrose \cite{Pen}, the
authors \cite{KP,Kspin}, Tom Etter \cite{Etter} and parts of our early paper on quantum knots \cite{QK1}.
\bigbreak

First, consider the multiplication of matrices $M = (M_{ij})$ and $N=(N_{kl})$ where $M$ is $m \times n$ and $N$ is $n \times p.$
Then $MN$ is $m \times p$ and $$(MN)_{ij} = \Sigma_{k=1}^{n} M_{ik}N_{kj}.$$  We represent each matrix by a box, and each index for 
the matrix elements by a line segment that is attached to this box. The common index in the summation is represented by a line that 
emanates from one box, and terminates in the other box. This line segment has no free ends. By the definition of matrix multipliction,
a line segment without free ends represents the summation over all possible index assignments that are available for that segment.
Segments with free ends correspond to the possible index choices for the product matrix. See Figure 10.
\bigbreak

\begin{center}
{\tt    \setlength{\unitlength}{0.92pt}
\begin{picture}(353,116)
\thinlines    \put(49,65){\framebox(41,41){}}
              \put(169,65){\framebox(41,40){}}
              \put(50,13){\framebox(41,41){}}
              \put(130,13){\framebox(41,41){}}
              \put(59,82){$M$}
              \put(180,82){$N$}
              \put(60,32){$M$}
              \put(140,32){$N$}
              \put(68,84){\line(0,0){0}}
              \put(130,32){\line(-1,0){40}}
              \put(50,33){\line(-1,0){40}}
              \put(170,33){\line(1,0){40}}
              \put(89,85){\line(1,0){22}}
              \put(146,86){\line(1,0){23}}
              \put(209,85){\line(1,0){21}}
              \put(49,86){\line(-1,0){20}}
              \put(229,29){$=$}
              \put(286,10){\framebox(40,42){}}
              \put(285,32){\line(-1,0){21}}
              \put(325,31){\line(1,0){18}}
              \put(292,28){$MN$}
\end{picture}}

{\bf Figure 10 -  Matrix Multiplication}
\end{center}

\noindent The trace of an $m \times m$ matrix $M$ is given by the formula $$tr(M) = \Sigma_{i=1}^{m} M_{ii}.$$
In diagrammatic terms the trace is represented by a box with the output segment identified with the input segment.
See Figure 11 for this interpretation of the matrix trace. 
\bigbreak

\begin{center}
{\tt    \setlength{\unitlength}{0.92pt}
\begin{picture}(186,92)
\thinlines    \put(113,48){$= tr(M)$}
              \put(53,29){$M$}
              \put(12,35){\line(0,1){45}}
              \put(12,35){\line(1,0){26}}
              \put(107,80){\line(-1,0){95}}
              \put(107,34){\line(0,1){46}}
              \put(90,34){\line(1,0){17}}
              \put(38,10){\framebox(53,44){}}
\end{picture}}

{\bf Figure 11 -  Matrix Trace}
\end{center}

\begin{center}
{\tt    \setlength{\unitlength}{0.92pt}
\begin{picture}(425,279)
\thinlines    \put(61,165){$tr(M)$}
\thicklines   \put(10,13){Result of Measurement}
              \put(138,198){Prepare}
              \put(145,217){\vector(1,0){33}}
\thinlines    \put(122,181){\line(-1,0){85}}
              \put(131,248){\line(0,-1){67}}
              \put(36,249){\line(0,-1){68}}
              \put(62,224){\framebox(53,44){}}
              \put(114,248){\line(1,0){17}}
              \put(36,249){\line(1,0){26}}
              \put(77,243){$M$}
              \put(131,181){\line(-1,0){12}}
              \put(197,155){$|\psi \rangle = M| b \rangle$}
              \put(297,182){\line(-1,0){12}}
              \put(202,181){\line(1,0){14}}
              \put(243,244){$M$}
              \put(202,250){\line(1,0){26}}
              \put(280,249){\line(1,0){17}}
              \put(228,225){\framebox(53,44){}}
              \put(202,250){\line(0,-1){68}}
              \put(297,249){\line(0,-1){67}}
              \put(282,198){\line(0,-1){30}}
              \put(282,198){\line(-3,-2){24}}
              \put(258,181){\line(2,-1){24}}
              \put(269,179){$b$}
              \put(99,46){$b$}
              \put(53,45){$a$}
              \put(88,48){\line(2,-1){24}}
              \put(112,65){\line(-3,-2){24}}
              \put(112,65){\line(0,-1){30}}
              \put(70,47){\line(-5,-3){23}}
              \put(47,65){\line(4,-3){23}}
              \put(47,64){\line(0,-1){30}}
              \put(127,116){\line(0,-1){67}}
              \put(32,117){\line(0,-1){68}}
              \put(58,92){\framebox(53,44){}}
              \put(110,116){\line(1,0){17}}
              \put(32,117){\line(1,0){26}}
              \put(73,111){$M$}
              \put(32,48){\line(1,0){14}}
              \put(127,49){\line(-1,0){12}}
              \put(135,107){$\rho_{ab} = | a \rangle \, \langle b |$}
              \put(135,79){$\langle a | M | b \rangle = tr(\rho_{ab} M)$}
\end{picture}}

{\bf Figure 12 -  Network Operations: Preparation and Measurement via Insertion of Bras and Kets.}
\end{center}

In Figure 12 we illustrate the diagrammatic interpretation of the formula
$$\langle a | M | b \rangle = tr(\rho_{ab} M).$$
This formula gives the amplitude for measuring the state $|a \rangle$ from a preparation of $|\psi \rangle = M | b \rangle.$
Note that the state $| \psi \rangle$ is obtained from the graphical structure of $tr(M)$ by cutting the connection between the 
input line and output line of the box labeled $M,$ and inserting the ket $| b \rangle$ on the output line. The resulting network
is shown at the top of Figure 12. This network, with one free end, represents the quantum state $| \psi \rangle.$ This state is the superposition
of all possible values (qubits) that can occur at the free end of the network. When we measure the state, one of the possible qubits occurs. 
The amplitude for the occurrence of $| a \rangle$ is equal to $\langle a | M | b \rangle.$ When we insert $| a \rangle$ at the free end of the network
for $| \psi \rangle,$ we obtain the network whose value is this amplitude.
\bigbreak 
 
\noindent If $M$ is unitary, then we can interpret the formula as the amplitude for measuring state $|a \rangle$ from 
a preparation in $| b \rangle,$ and an evolution of this preparation by the unitary transformation $M.$ In the first
interpretation the operator $M$ can be an observable, aiding in the preparation of the state. In this notation,
$$\rho_{ab} = | a \rangle \, \langle \, b \, |$$ is the ket-bra associated with the states $| a \rangle$ and $| b \rangle.$
If $a=b$, then $\rho_{aa}$ is the density matrix associated with the pure state $| a \rangle.$ 
\bigbreak

The key to this graphical model for preparation and measurement is the understanding that the diagram is both a combinatorial structure {\it and} a representative
for the computation of either an amplitude or a state (via summation over the indices available for the internal lines and superposition over the possibilities for
the free ends of the network). A diagram with free ends (no kets or bras tied into the ends) represents a state that is the superposition of all the possibilities
for the values of the free ends. This superposition is a superposition of diagrams with different labels on the ends. In this way the principles of quantum
measurement are seen to live in categories of diagrams. A given diagram can be regarded as a world that is subject to preparation and measurement. After such
an operation is performed (Cut an edge. Insert a density matrix.), a new world is formed that is itself subject to preparation and measurement. This succession 
of worlds and states can be regarded as a description of the evolution of a quantum process. 
\bigbreak

\noindent {\bf Remark.} One can generalize this notion of quantum process in networks by allowing the insertion of other operators into the network, and by
allowing systematic operations on the graph. Techniques of this sort are used in spin foam models for quantum gravity \cite{M}, and in renormalization of
statistical mechanics models. The reader may also find it fruitful to compare the approaches in the present paper with the work in \cite{KnotAutomata,Lambda1,Lambda2} where we work on the relationship of classical computation and network structures.
\bigbreak

 \end{document}